\documentstyle[aps,prl,epsfig]{revtex}
\tightenlines

\newcommand{\be}{\begin{equation}}
\newcommand{\ee}{\end{equation}}

\newcommand{\bmath}{\begin{mathletters}}
\newcommand{\emath}{\end{mathletters}}

\begin{document}

\title{\Large{\bf Antiferromagnetic ordering of itinerant systems in modified mean-field theory}}

\vskip0.5cm 

\author{ G. G\'{o}rski, J. Mizia, K. Kucab }

\address{Institute of Physics, University of Rzesz\'{o}w, ulica Rejtana 16A, \\
35-958 Rzesz\'{o}w, Poland\\}

\maketitle

\vskip0.5cm 


\begin{abstract}

\noindent
 This is an analysis of the itinerant model for antiferromagnetism, in which is included both on-site and inter-site electron correlations. We also consider the band degeneration, 
which brings into the Hamiltonian the on-site exchange interactions. 

\noindent
The Green function technique is used and the coherent potential approximation (CPA) decoupling for the on-site Coulomb repulsion. This decoupling combined with the modified 
Hartree-Fock approximation for the inter-site interactions generates the change in shape of the spin bands with growing interaction constants, which is described by the 
correlation factors and which decreases the kinetic energy of the system. The effective Hartree field and the gain in kinetic energy due to the on-site and inter-site correlation 
factors drive the antiferromagnetism. The on-site and inter-site interactions act together towards the antiferromagnetism. Their cooperation decreases the interaction constants 
required for the antiferromagnetic ordering.

\noindent
This new approach allows for the antiferromagnetic instability in the purely itinerant model at the half-filling in the split band limit. This situation describes the high temperature 
superconducting cuprates.

\end{abstract}
\vskip0.5cm



\vskip1.5cm

\noindent {\Large {\bf 1. Introduction} }
\vskip0.5cm

The basic model for magnetic ordering of itinerant electrons in solids is the Hubbard model \cite{1}. The largest interaction between itinerant electrons is the on-site Coulomb 
repulsion, $U=(i,i\mid 1/r\mid i,i)$ , where $i$ is the lattice site index. In the mean-field approximation the Hubbard model leads to the well-known Stoner model for magnetism.

Extension of the Hubbard model is the model in which in addition to the on-site Coulomb repulsion $U$, there is also the Coulomb exchange on-site (between different orbitals) 
interaction $J_{in}$. The magnetic solutions of the three-dimensional Hubbard model did not make the exchange field necessary for antiferromagnetism to be small enough to be 
justified physically (see e.g. Ref. \cite{2}), therefore the authors failed to arrive at the magnetic state. More recently Hirsch \cite{3,4} and others (see e.g. Refs \cite{5,6,7}) also 
included the nearest-neighbor inter-site interactions; $(i,j\mid 1/r\mid k,l)$, where $(k,l)=(i,j)$ and $i,j$ are the nearest-neighbor lattice sites.

In our degenerate band model, we introduced into the Hamiltonian the on-site Coulomb repulsion \mbox{$U=(i\lambda,i\lambda \mid 1/r\mid i\lambda,i\lambda)$}, the on-site two 
sub-band interactions $\left( {i\lambda ,i\nu |1/r|i\vartheta ,i\varpi } \right)$ (where the sub-band indices $\left( {\varpi ,\vartheta } \right) = \left( {\lambda ,\nu } \right)$ and 
$\lambda  \ne \nu$), as well as the inter-site nearest-neighbor interactions $(i,j\mid 1/r\mid k,l)$ (for $(k,l)=(i,j)$ and $i \ne j$).
\widetext
Through analyzing the inter-site correlations, the product of four operators is dealt with; two of which are replaced by its stochastic value. The stochastic value of the 
two-operator product has been interpreted as the product of probabilities for electron transfer between two atoms. This interpretation explains that Hirsch's \cite{4} the average 
bond occupation for spin $\sigma$; $i_\sigma$, comes from the standard probabilities used in the coherent potential approximation (CPA) method \cite{8} when it is applied to the 
inter-site correlation in the lowest mean-field (or Hartree-Fock) approximation. This connection will allow the expansion of the model in the future to more realistic cases of 
inter-site interactions of arbitrary strength. In this paper the weak inter-site interaction is accompanied by the weak or strong on-site Coulomb interaction. In the latter case the 
self-energy has higher order expansion in powers of interaction constant over the bandwidth. The spin band narrowing (broadening) is arising from inter-site interactions (see 
Refs \cite{3,9}).

The paper is organized as follows. In Section 2, the model Hamiltonian is put forward and in Section 3 the CPA formalism is developed to treat the  inter-site Coulomb correlation. 
In Section 4, we set up the model for antiferromagnetism, which includes the Coulomb on-site and inter-site correlations. Numerical examples are presented in Section 5 based on 
the constant density of state (DOS) for the weak inter-site interactions and weak or strong on-site Coulomb interaction. Finally, Section 6 is devoted to the conclusions.

\vskip 0.5cm 

\noindent {\Large {\bf 2. The model}} 

\vskip0.5cm 

The Hamiltonian for the single degenerate band can be written in the form which has been introduced previously (see Ref.  \cite{10}) 
\samepage
\begin{eqnarray}
  H &=&  - t_0 \sum\limits_{ < ij > \sigma } {\left( {c_{i\sigma }^ +  c_{j\sigma }  + h.c.} \right)}  - \mu _0 \sum\limits_i {\hat n_i }  - F_{ \rm in}^0 \sum\limits_{i,\sigma } 
{n_i^\sigma  \hat n_{i\sigma } }  + U\sum\limits_{i\sigma } {\hat n_{i\sigma } \hat n_{i - \sigma } } \nonumber \\
& +& J\sum\limits_{ < ij > \sigma \sigma '} {c_{i\sigma }^ +  c_{j\sigma '}^ +  c_{i\sigma '} c_{j\sigma } }  
   + V\sum\limits_{ < ij > } {\hat n_i \hat n_j }  + J'\sum\limits_{ < ij > \sigma } {c_{i\sigma }^ +  c_{i - \sigma }^ +  c_{j - \sigma } c_{j\sigma } }\nonumber \\
   &+& \sum\limits_{ < ij > \sigma } {\Delta t\left( {c_{i\sigma }^ +  c_{j\sigma }  + h.c.} \right)\left( {\hat n_{i - \sigma }  + \hat n_{j - \sigma } } \right)}  
\label{1}
\end{eqnarray}
where $t_0$ is the nearest-neighbors hopping integral, $\mu _0$ is the chemical potential, $c_{i\sigma }^ +  \left( {c_{i\sigma } } \right)$ creates (destroys) an electron of spin 
$\sigma$ on the $i$th lattice site, $\hat n_{i\sigma }  = c_{i\sigma }^ +  c_{i\sigma }$ is the electron number operator for electrons with spin $\sigma$ on the $i$th lattice site, $\hat 
n_i  = \hat n_{i\sigma }  + \hat n_{i - \sigma }$ is the operator of the total number of electrons on the $i$th lattice site, $n_i^\sigma$ is the average number of electrons on sites $i$ 
with spin $\sigma$. The model Hamiltonian (\ref{1}) includes explicitly the band degeneration already in the Hartree-Fock approximation. The internal field constant $F_{\rm 
in}^0$ is given by
\be
F_{\rm in}^0  = (p - 1)(J_{\rm in}  + J'_{\rm in}  + V_{\rm in} ),
\label{2}
\ee
where $p$ is the number of sub-bands in the band and $J_{\rm in}=(i\lambda,i\nu \mid 1/r\mid i\nu,i\lambda)$, $J'_{\rm in}=(i\lambda,i\lambda \mid 1/r\mid i\nu,i\nu)$, $V_{\rm 
in}=(i\lambda,i\nu \mid 1/r\mid i\lambda,i\nu)$ are the on-site two sub-band inter orbital interactions of the following types: exchange, pair hopping and density-density, 
respectively (see Ref. \cite{10}). The indices $\lambda, \nu$ numerate the sub-bands in the degenerate single band.

The remaining potential part of the Hamiltonian contains the on-site Coulomb repulsion $U=(i\lambda,i\lambda \mid 1/r\mid i\lambda,i\lambda)$, the inter-site exchange interaction 
$J=(i,j \mid 1/r\mid j,i)$, density-density interaction $V=(i,j \mid 1/r\mid i,j)$, pair-hopping interaction $J'=(i,i \mid 1/r\mid j,j)$ and assisted hopping interaction $\Delta t=(i,i \mid 
1/r\mid j,i)$.

The on-site Coulomb repulsion will be set aside for the time being and only inter-site terms will be consider in the Hamiltonian (\ref{1}). The following form is arrived at
\begin{eqnarray}
H &=& \sum\limits_{k\sigma } {(\varepsilon _k^0  - \mu _0 )\hat n_{k\sigma } }  - F_{\rm in}^0 \sum\limits_{i\sigma } {n_i^\sigma  \hat n_{i\sigma } }  + \sum\limits_{ < ij > 
\sigma } {\Delta t(\hat n_{i - \sigma }  + \hat n_{j - \sigma } )\left( {c_{i\sigma }^ +  c_{j\sigma }  + c_{j\sigma }^ +  c_{i\sigma } } \right)} \nonumber \\
&+&\sum\limits_{ < ij > \sigma } {\left[ {Jc_{j - \sigma }^ +  c_{i - \sigma }  + J'c_{i - \sigma }^ +  c_{j - \sigma }  + (J - V)c_{j\sigma }^ +  c_{i\sigma } } \right]c_{i\sigma }^ +  
c_{j\sigma } }, 
\label{3}
\end{eqnarray}
where $\hat n_{k\sigma }  = c_{k\sigma }^ +  c_{k\sigma }$, $c_{k\sigma }^ +  \left( {c_{k\sigma } } \right)$ creates (destroys) an electron of spin $\sigma$ and momentum $k$ 
and $\varepsilon _k^0$ is the unperturbed dispersion relation for electron
\be
\varepsilon _k^0  =  - t_0 \gamma _k,
\;\;\;\;
\gamma _k  = \sum\limits_{ < i,j > } {e^{ik\left( {R_i  - R_j } \right)} }.
\label{4}
\ee
Next we will analyse Hamiltonian (\ref{3}) using the coherent potential (CP) approximation scheme and its lowest order mean field decoupling.

\vskip1.0cm 
\noindent {\Large {\bf 3.  CP approximation for inter-site interactions}} 
\vskip0.5cm 

The CPA idea is now applied to the inter-site weak interactions; $J, J', V$ and $\Delta t$, in Hamiltonian (\ref{3}) and the stochastic value will replace each operator product in the 
square bracket. For example, the operator product; $c_{j\sigma }^ +  c_{i\sigma }$ will be replaced by its stochastic value $\overline {c_{j\sigma }^ +  c_{i\sigma } }$. It takes the 
values 1 or 0. The probability of value 1 is the probability of electron with spin $\sigma$ hopping from the $i$ to the $j$ lattice site and back. This probability, denoted by 
$I_\sigma$ below, is given by the average of two products. One is the product of probabilities that there is an electron with spin $\sigma$ on the $i$ site and that the $j$ site has 
empty states; $n_i^\sigma  (n_j^{t\sigma }  - n_j^\sigma  )/\left( {n_i^\sigma   + (n_j^{t\sigma }  - n_j^\sigma  )} \right)$, and the second one is the probability of the opposite 
jump; $n_j^\sigma  (n_i^{t\sigma }  - n_i^\sigma  )/\left( {n_j^\sigma   + (n_i^{t\sigma }  - n_i^\sigma  )} \right)$. The quantity $n_i^\sigma$ is the average number of electrons on 
sites $i$ with spin $\sigma$ and $n_i^{t\sigma }$ is the total capacity of the sub-band $\sigma$ on site $i$. The total hopping probability, called $I_\sigma$, is given by the 
average of above probabilities
\be
I_\sigma   \equiv \frac{1}{2}\left[ {\frac{{n_i^\sigma  (n_j^{t\sigma }  - n_j^\sigma  )}}{{n_i^\sigma   + (n_j^{t\sigma }  - n_j^\sigma  )}} + \frac{{n_j^\sigma  (n_i^{t\sigma }  - 
n_i^\sigma  )}}{{n_j^\sigma   + (n_i^{t\sigma }  - n_i^\sigma  )}}} \right].
\label{5}
\ee
For the weak correlation, when the band is not split by the Coulomb repulsion, we have $n_i^{t\sigma }=1$ and
\be
I_\sigma   \equiv \frac{1}{2}\left[ {\frac{{n_i^\sigma  (1 - n_j^\sigma  )}}{{n_i^\sigma   + (1 - n_j^\sigma  )}} + \frac{{n_j^\sigma  (1 - n_i^\sigma  )}}{{n_j^\sigma   + (1 - 
n_i^\sigma  )}}} \right].
\label{6}
\ee
This expression for the ferromagnetism $\left( {n_i^\sigma   = n_j^\sigma   \equiv n^\sigma  } \right)$ can be simplified to\footnote{Micnas et al. \cite{9} and Hirsch \cite{4} define 
the average band occupation for spin $\sigma$ as $i_\sigma   = \int\limits_{ - D_0 }^{D_0 } {\rho ^\sigma  \left( { - \frac{\varepsilon }{{D_0 }}} \right)f_\sigma  (\varepsilon 
)d\varepsilon }$. For the constant DOS this will give $i_\sigma   = n_i^\sigma  (1 - n_j^\sigma  )$ which is identical to our $I_\sigma$ from Eq. (\ref{7}).}
\be
i_\sigma   = n_i^\sigma  (1 - n_j^\sigma  ).
\label{7}
\ee
In the case of strong correlation when the band is split into the lower and upper Hubbard sub-band, $n_i^{t\sigma}$ is the total capacity of the sub-band in which the chemical 
potential is located.  For example, for the lower Hubbard sub-band we have; $n_i^{t\sigma }  = 1 - n_i^{ - \sigma }$. These values of $n_i^{t\sigma}$ together with the definitions 
of $n_i^\sigma$ for antiferromagnetism will be used later to calculate $I_\sigma   = I_{ - \sigma }  \equiv I_{AF}$ in the case of antiferromagnetism.

After adding up and subtracting the inter-site self-energy $\Sigma _{1,2}^\sigma$ into the Hamiltonian (\ref{3}) and performing the above-described stochastic approximation, 
this Hamiltonian takes on the following form:
\be
H = \sum\limits_{k\sigma } {(\varepsilon _k^\sigma   - \mu _0 )\hat n_{k\sigma } }  + \sum\limits_{\scriptstyle \left\langle {i,j} \right\rangle  \hfill \atop 
  \scriptstyle \;\;\sigma  \hfill} {\left( {\tilde \varepsilon  - \Sigma _{1,2}^\sigma  } \right)(c_{i\sigma }^ +  c_{j\sigma }  + c_{j\sigma }^ +  c_{i\sigma } )}  + \sum\limits_{i\sigma } 
{\left( {\tilde M_i^\sigma   - F_{\rm in}^0 n_i^\sigma  } \right)\hat n_{i\sigma } }, 
\label{8}
\ee
where the effective dispersion relation is given by
\be
\varepsilon _k^\sigma   = \varepsilon _k^0  + \gamma _k \Sigma _{1,2}^\sigma   = \varepsilon _k^0 L_\sigma. 
\label{9}
\ee
The quantity $L_\sigma$ is the parameter of the bandwidth change taking the following form:
\be
L_\sigma   = \frac{{D_\sigma  }}{{D_0 }} = 1 - \frac{{\Sigma _{1,2}^\sigma  }}{{t_0 }},
\label{10}
\ee
where $D_0=zt_0$ is the unperturbed half bandwidth, $z$ is the number of nearest neighbors and $D_\sigma$ is  the spin-dependent effective (perturbed) half bandwidth, for 
which we can write $D_\sigma=zt_\sigma$, where the perturbed hopping integral $t_\sigma=L_\sigma t_0$.

The quantity $\tilde M_i^\sigma$ in Eq. (\ref{8}) is the stochastic energy describing the inter-site interactions
\be
\tilde M_i^\sigma   =  - J\sum\limits_j {'\bar n_{j\sigma } }  + V\sum\limits_j {'(\bar n_{j - \sigma }  + \bar n_{j\sigma } )}  + 2\Delta t\sum\limits_j {'\overline {c_{i - \sigma }^ +  
c_{j - \sigma } } },
\label{11}
\ee
where $\sum\nolimits_j^{'} {}$ is the sum over the nearest neighbors of the lattice site $i$, the bar over the operators product defines this product as the stochastic variable.

The stochastic energy $\tilde M_i^\sigma$ together with the on-site Coulomb repulsion can be subsequently analyzed in the CP approximation (see Ref. \cite{8}).

The stochastic potential in Hamiltonian (\ref{8}) can be expressed as
\be
\tilde \varepsilon  = (J + J')\overline {c_{j - \sigma }^ +  c_{i - \sigma } }  + (J - V)\overline {c_{j\sigma }^ +  c_{i\sigma } }  + \Delta t\left( {\overline {n_{i - \sigma } }  + \overline 
{n_{j - \sigma } } } \right),
\label{12}
\ee
with the stochastic values 0 and 1 for each operator product in Eq. (\ref{12}). The stochastic potential $\tilde \varepsilon$ will take on 16 different values, $\varepsilon _l$, with 
corresponding probabilities, $P_l^\sigma$, given by Eq. (A2) in Appendix. For these probabilities and potentials it is now possible to write the following equation for the 
self-energy $\Sigma _{1,2}^\sigma  $, which describes only the inter-site interactions in Hamiltonian (\ref{8})
\be
\sum\limits_{l = 1}^{16} {P_l^\sigma  \frac{{\varepsilon _l  - \Sigma _{1,2}^\sigma  }}{{1 - \left( {\varepsilon _l  - \Sigma _{1,2} } \right)F^\sigma  \left( \varepsilon  \right)}}}  = 0,
\label{13}
\ee
where the Slater-Koster function ${F^\sigma  \left( \varepsilon  \right)}$ has the following form
\be
F^\sigma  \left( \varepsilon  \right) = \frac{1}{N}\sum\limits_k {\frac{1}{{\varepsilon  + \mu _0  - \varepsilon _k^0  - \gamma _k \Sigma _{1,2}^\sigma  }}}  = 
\frac{1}{N}\sum\limits_k {\frac{1}{{\varepsilon  + \mu _0  - \varepsilon _k^0 L_\sigma  }}} 
\label{14}
\ee
The Slater-Koster function can be written as
\be
F^\sigma  \left( \varepsilon  \right) = \frac{1}{N}\sum\limits_k {G^\sigma  (\varepsilon ,k)}, 
\label{15}
\ee
where the Green function ${G^\sigma  (\varepsilon ,k)}$ is given by
\be
G^\sigma  (\varepsilon ,k) = \frac{1}{{\varepsilon  + \mu _0  - \varepsilon _k^0 L_\sigma  }} = \frac{1}{{\varepsilon  + \mu _0  - \varepsilon _k^\sigma  }}.
\label{16}
\ee
In the first order approximation, for the weak inter-site interaction, we obtain the following formula for the inter-site self-energy $\Sigma _{1,2}^\sigma$
\be
\Sigma _{1,2}^\sigma   \cong \sum\limits_{l = 1}^{16} {\varepsilon _l P_l^\sigma  }  = (J - V)I_\sigma   + (J + J')I_{ - \sigma }  + \Delta t(n_i^{ - \sigma }  + n_j^{ - \sigma } )
\label{17}
\ee
and the parameter of the bandwidth change in Eq. (\ref{16}) is given by
\be
L_\sigma   = 1 - \frac{{\Sigma _{1,2}^\sigma  }}{{t_0 }} = 1 - \frac{{\Delta t}}{{t_0 }}(n_i^{ - \sigma }  + n_j^{ - \sigma } ) - \frac{{\left( {J + J'} \right)}}{{t_0 }}I_{ - \sigma }  - 
\frac{{\left( {J - V} \right)}}{{t_0 }}I_\sigma,
\label{18}
\ee
where $n_i^{ \sigma }$ is the average number of electrons with spin $\sigma$ on the $i$th lattice site.

In the case of ferromagnetism, the parameter of the bandwidth change $L_\sigma$ was spin dependent (see e.g. Refs \cite{3,10}) which caused the spin dependence of the 
hopping integral $t_\sigma$. The value of the hopping integral for the majority spin $\sigma$ decreased with respect to the hopping integral for spin $-\sigma$.

For the antiferromagnetism the indices $i, j$ belong to the neighboring sub-lattices $\alpha, \beta$ with opposite magnetic moments. In the result parameter; 
$I_\sigma=I_{-\sigma}\equiv I_{\rm AF}$ is spin independent (i.e. it does not depend on the first power of antiferromagnetic moment $m$) and the bandwidth factor from Eq. 
(\ref{18})
\be
L_{\rm AF}  = \frac{D}{{D_0 }} = 1 - n\frac{{\Delta t}}{{t_0 }} + \frac{1}{{t_0 }}\left( { - 2J - J' + V} \right)I_{\rm AF} 
\label{19}
\ee
is also spin independent (here $n$ is the total number of electrons).

The factor $L_{\rm AF}$ is spin independent and because of it the dispersion relation will also not depend on the spin direction $\varepsilon _k^\sigma   = \varepsilon _k^{ - 
\sigma }  \equiv \varepsilon _k  = \varepsilon _k^0 L_{\rm AF}$. The Green function from Eq. (\ref{16}) will now take on the form
\be
G(\varepsilon ',k) = \frac{1}{{\varepsilon ' - \varepsilon _k }}, \;\;\;\; \varepsilon ' = \varepsilon  + \mu _0,
\label{20}
\ee
which will be used in the next section in the case of antiferromagnetic ordering.

In the first order approximation the stochastic inter-site energy $\tilde M_i^\sigma$ given by Eq. (\ref{11}) becomes the molecular field, which is given by the expression:
\be
M_i^\sigma   =  - J\sum\limits_j {'n_j^\sigma  }  + zVn + z\Delta tI_{\rm AF}. 
\label{21}
\ee 
In further analysis of the antiferromagnetism, the first order approximation for the parameter of the bandwidth change, $L_{\rm AF}$ will be used (Eq. (\ref{19})) and the 
stochastic energy $\tilde M_i^\sigma$ will be given by Eq. (\ref{21}). 

\newpage
\noindent {\Large {\bf 4. Antiferromagnetism (AF)}} 
\vskip0.5cm 

For the antiferromagnetism, the diagonalization of Plischke and Mattis \cite{11}, Brouers \cite{12} and Mizia \cite{13} will be used. The crystal lattice will be divided into two 
interpenetrating sub-lattices $\alpha, \beta$, with the average electron numbers equal to
\be
n_\alpha ^{ \pm \sigma }  = \frac{{n \pm m}}{2}, \;\;\;\; n_\beta ^{ \pm \sigma }  = \frac{{n \mp m}}{2},
\label{22}
\ee
where $m$ is the antiferromagnetic moment per atom in Bohr's magnetons.

The reduced part of the Hamiltonian (\ref{8}) together with the on-site Coulomb correlation is given by
\begin{eqnarray}
H &=&  - t\sum\limits_{\scriptstyle i,j  \hfill \atop 
  \scriptstyle \,\sigma  \hfill} {\left( {\alpha _{i\sigma }^ +  \beta _{j\sigma }  + \beta _{j\sigma }^ +  \alpha _{i\sigma } } \right)}  - \mu _0 \sum\limits_{\scriptstyle i,\gamma,\sigma  
\atop (\gamma  = \alpha ,\beta ) } {\hat n_{i\sigma }^\gamma  }  \nonumber \\
  &+& \sum\limits_{\scriptstyle i,\gamma,\sigma \atop 
  (\gamma  = \alpha ,\beta ) \hfill} {\left( {M_{i,\gamma }^\sigma   - F_{in}^0 n_\gamma ^\sigma  } \right)\hat n_{i\sigma }^\gamma  }  + U\sum\limits_{\scriptstyle i,\gamma,\sigma 
\atop (\gamma  = \alpha ,\beta ) \hfill} {\hat n_{i\sigma }^\gamma  \hat n_{i, - \sigma }^\gamma  } 
\label{23}
\end{eqnarray}
where $\alpha _{i\sigma }^ +  (\alpha _{i\sigma } )$ and $\beta _{i\sigma }^ +  (\beta _{i\sigma } )$ are the creation (annihilation) operators for an electron of spin $\sigma$ on 
sub-lattice $\alpha$ and $\beta$ respectively, $\hat n_{i\sigma }^\alpha   = \alpha _{i\sigma }^ +  \alpha _{i\sigma }$ is the electron number operator for electrons with spin 
$\sigma$ on the sub-lattice $\alpha$, and $t=t_0L_{\rm AF}$ is the effective hopping integral. The exchange potential in the mean-field approximation is different than in the case 
of ferromagnetism since the moments on neighboring lattice sites are opposite to each other and in effect both the inter-site and on-site component of the exchange potential will 
depend on the type of lattice site ($\alpha$ or $\beta$). Combining Eq. (\ref{21}) with Eq. (\ref{22}) we obtain for the inter-site component that
\be
M_{i,\alpha \left( \beta  \right)}^\sigma   \equiv M_{\alpha \left( \beta  \right)}^\sigma   =  - Jzn_{\beta \left( \alpha  \right)}^\sigma   + zVn + z\Delta tI_{\rm AF}. 
\label{24}
\ee
Now we will derive the equations for dispersion relation, particle number and magnetization for the antiferromagnetic state using Hamiltonian (\ref{23}) and the Green function 
technique. Mean-field approximation is used for the parameter of the bandwidth change $L_{\rm AF}$ and for the stochastic energy $\tilde M_i^\sigma   \equiv M_{\alpha \left( 
\beta  \right)}^\sigma$ (Eq. (\ref{24})). This approximation was performed directly on the Hamiltonian, as there is no difference with the case when it is done within the Green 
function chain of equations. For the on-site interaction, $U$, we use the standard CPA, which for simplicity, in further analysis, is treated in the weak and strong correlation limit. 
For the weak correlation, $(U\ll D)$, we use first order approximation in interaction constant over the bandwidth. This is equivalent to the Hartree-Fock or virtual crystal 
approximation. The second case is the high correlation, $(U\gg D)$, approximation. This approach will allow for easy extension of the calculations to the arbitrary strength of the 
on-site interaction later on.

The main idea of the CPA formalism \cite{8} is used now. We split the above stochastic Hamiltonian (\ref{23}) into a homogeneous part
\be
H_0  =  - t\sum\limits_{\scriptstyle i,j\atop \sigma } {\left( {\alpha _{i\sigma }^ +  \beta _{j\sigma }  + \beta _{j\sigma }^ +  \alpha _{i\sigma } } \right)}  - \mu 
\sum\limits_{\scriptstyle i,\gamma, \sigma \atop (\gamma  = \alpha ,\beta )} {\hat n_{i\sigma }^\gamma  }  + \sum\limits_{i\sigma } {\Sigma _\alpha ^\sigma  \hat n_{i\sigma 
}^\alpha  }  + \sum\limits_{i\sigma } {\Sigma _\beta ^\sigma  \hat n_{i\sigma }^\beta  } 
\label{25}
\ee
and a stochastic part
\be
H_I  = \sum\limits_{i\sigma } {(\tilde V_\alpha ^\sigma   - \Sigma _\alpha ^\sigma  )\hat n_{i\sigma }^\alpha  }  + \sum\limits_{i\sigma } {(\tilde V_\beta ^\sigma   - \Sigma _\beta 
^\sigma  )\hat n_{i\sigma }^\beta  }, 
\label{26}
\ee
where
\be
\mu  = \mu _0  - zVn - 2z\Delta tI_{\rm AF} 
\label{27}
\ee
is the effective chemical potential, $\Sigma _\gamma ^\sigma$ are the self energies on sites $\gamma=\alpha(\beta)$ for electrons with spin $\sigma$ and $\tilde V_{\alpha (\beta 
)}^\sigma$ are the stochastic potentials given by
\[
\tilde V_{\alpha (\beta )}^\sigma   = \left\{ \begin{array}{l}
 \tilde V_{1\alpha (\beta )}^\sigma   =  - F_{\rm in}^0 n_{\alpha (\beta )}^\sigma   - zJn_{\beta (\alpha )}^\sigma   \\ 
 \tilde V_{2\alpha (\beta )}^\sigma   = U - F_{\rm in}^0 n_{\alpha (\beta )}^\sigma   - zJn_{\beta (\alpha )}^\sigma   \\ 
 \end{array} \right.,
\]
with probabilities
\be
\begin{array}{l}
 P_{1\alpha (\beta )}^\sigma   = 1 - n_{\alpha (\beta )}^{ - \sigma }  \\ 
 P_{2\alpha (\beta )}^\sigma   = n_{\alpha (\beta )}^{ - \sigma }  \\ 
 \end{array}.
\label{28}
\ee
The self-energies $\Sigma _\gamma ^\sigma$ fulfill the equations
\be
\sum\limits_{i = 1}^2 {P_{i\gamma }^\sigma  \frac{{\tilde V_{i\gamma }^\sigma   - \Sigma _\gamma ^\sigma  }}{{1 - \left( {\tilde V_{i\gamma }^\sigma   - \Sigma _\gamma ^\sigma  
} \right)F_\sigma ^{\gamma \gamma } (\varepsilon )}}}  = 0, \;\;\;\; \gamma  = \alpha (\beta ),
\label{29}
\ee
with the Slater-Koster function $F_\sigma ^{\gamma \gamma } (\varepsilon )$ in the following form
\be
F_\sigma ^{\gamma \gamma } (\varepsilon ) = \frac{1}{N}\sum\limits_k {G_\sigma ^{\gamma \gamma } (\varepsilon ,k)} 
\label{30}
\ee
and ${G_\sigma ^{\gamma \gamma } (\varepsilon ,k)}$ given by Eq. (\ref{35}) below.
We transform Hamiltonian (\ref{25}) into momentum space, and use it in the equations of motion for the Green functions:
\be
\varepsilon \left\langle {\left\langle {A;B} \right\rangle } \right\rangle _\varepsilon   = \left\langle {\left[ {A,B} \right]_ +  } \right\rangle  + \left\langle {\left\langle {\left[ {A,H_0 } 
\right]_ -  ;B} \right\rangle } \right\rangle _\varepsilon,  
\label{31}
\ee 
where $(A,B) \in \left( {\alpha _{k\sigma }^ +  ,\alpha _{k\sigma } ,\beta _{k\sigma }^ +  ,\beta _{k\sigma } } \right)$. The following equations are obtained
\be
\left[ {\begin{array}{*{20}c}
   {\varepsilon  + \mu  - \Sigma _\alpha ^\sigma  } & { - \varepsilon _k }  \\
   { - \varepsilon _k } & {\varepsilon  + \mu  - \Sigma _\beta ^\sigma  }  \\
\end{array}} \right]\left[ {\begin{array}{*{20}c}
   {G_\sigma ^{\alpha \alpha } (\varepsilon ,k)} & {G_\sigma ^{\alpha \beta } (\varepsilon ,k)}  \\
   {G_\sigma ^{\beta \alpha } (\varepsilon ,k)} & {G_\sigma ^{\beta \beta } (\varepsilon ,k)}  \\
\end{array}} \right] = \left[ {\begin{array}{*{20}c}
   1 & 0  \\
   0 & 1  \\
\end{array}} \right],
\label{32}
\ee
with e.g. $G_\sigma ^{\alpha \alpha } (\varepsilon ,k) = \left\langle {\left\langle {\alpha _{k\sigma } ;\alpha _{k\sigma }^ +  } \right\rangle } \right\rangle$, $G_\sigma ^{\beta 
\alpha } (\varepsilon ,k) = \left\langle {\left\langle {\beta _{k\sigma } ;\alpha _{k\sigma }^ +  } \right\rangle } \right\rangle$.

Solving the set of these equations we arrive at the following expressions for the Green functions:
\be
\left[ {\begin{array}{*{20}c}
   {G_\sigma ^{\alpha \alpha } (\varepsilon ,k)} & {G_\sigma ^{\alpha \beta } (\varepsilon ,k)}  \\
   {G_\sigma ^{\beta \alpha } (\varepsilon ,k)} & {G_\sigma ^{\beta \beta } (\varepsilon ,k)}  \\
\end{array}} \right] = \frac{1}{{\rm Det}}\left[ {\begin{array}{*{20}c}
   {\varepsilon  + \mu  - \Sigma _\beta ^\sigma  } & {\varepsilon _k }  \\
   {\varepsilon _k } & {\varepsilon  + \mu  - \Sigma _\alpha ^\sigma  }  \\
\end{array}} \right],
\label{33}
\ee
where
\be
\rm Det = \left( {\varepsilon  + \mu  - \Sigma _\beta ^\sigma  } \right)\left( {\varepsilon  + \mu  - \Sigma _\alpha ^\sigma  } \right) - \varepsilon _k^2. 
\label{34}
\ee
From Eq. (\ref{33}) we have for $\gamma=\alpha(\beta)$
\be
G_\sigma ^{\gamma \gamma } (\varepsilon ,k) = \frac{1}{2}\sqrt {\frac{{\varepsilon  + \mu  - \Sigma _{\beta (\alpha )}^\sigma  }}{{\varepsilon  + \mu  - \Sigma _{\alpha (\beta 
)}^\sigma  }}} \left[ {G\left( {\varepsilon _{\rm eff} ,k} \right) - G\left( { - \varepsilon _{\rm eff} ,k} \right)} \right],
\label{35}
\ee
with
\be
\varepsilon _{\rm eff}  = \sqrt {\left( {\varepsilon  + \mu  - \Sigma _\alpha ^\sigma  } \right)\left( {\varepsilon  + \mu  - \Sigma _\beta ^\sigma  } \right)} 
\label{36}
\ee
and $G(\varepsilon _{\rm eff} ,k)$ given by Eq. (\ref{20}) with $\varepsilon'$ replaced by $\varepsilon _{\rm eff}$ and $\mu _0$ by $\mu$ from Eq. (\ref{27}):
\be
G\left( {\varepsilon _{\rm eff} ,k} \right) = \frac{1}{{\varepsilon _{\rm eff}  - \varepsilon _k }}, \;\;\;\; \varepsilon _k  = \varepsilon _k^0 L_{\rm AF}.
\label{37}
\ee
This is how way we take into the account the inter-site correlations described in the proceeding section.
 
Next we use these Green functions in the expressions for electron numbers
\be
n_\gamma ^{ \pm \sigma }  = \frac{1}{N}\sum\limits_k {\int\limits_{ - \infty }^\infty  {f\left( \varepsilon  \right)\left( { - \frac{1}{\pi }} \right)} {\mathop{\rm Im}\nolimits} \left[ 
{G_{ \pm \sigma }^{\gamma \gamma } \left( {\varepsilon ,k} \right)} \right]} d\varepsilon, 
\label{38}
\ee
where
\[
f\left( \varepsilon  \right) = \frac{1}{{1 + \exp [\varepsilon /k_B T]}}.
\]
Treating the self-energies as $\Sigma _\gamma ^\sigma$, which are solutions of Eq. (\ref{29}), and using it in the first order approximation we can obtain
\be
\Sigma _{\alpha \left( \beta  \right)}^\sigma   \cong \sum\limits_{i = 1}^2 {P_{i\alpha \left( \beta  \right)}^\sigma  \tilde V_{i\alpha \left( \beta  \right)}^\sigma  }  = \left( {U - F_{\rm 
in}^0  - zJ} \right)\frac{n}{2} \mp \sigma \left( { \pm \sigma } \right)\Delta, 
\label{39}
\ee
where the antiferromagnetic energy gap $\Delta$ is given by the expression
\be
\Delta  = F_{\rm AF} \frac{m}{2}
\label{40}
\ee
and $F_{\rm AF}$ is the value of the effective exchange interaction given by
\be
F_{\rm AF}  = F_{\rm in}  - zJ,
\label{41}
\ee
with $F_{\rm in}$, the on-site exchange interaction (between different orbitals), being originally $F^0_{\rm in}$ given by Eq. (\ref{2}). In the present case of a weak Coulomb 
correlation $U\left( {U/D \ll 1} \right)$, the on-site interaction, $F_{\rm in}$, is augmented in the derivation process of Eq. (\ref{40}), by the Coulomb interaction treated in the 
Hartree-Fock approximation. As a result, we have
\be
F_{\rm in}  = \left\{ \begin{array}{l}
 U + F_{\rm in}^0 {\;\;\;\;\rm for \;\;\;\;}{U \mathord{\left/
 {\vphantom {U D}} \right.
 \kern-\nulldelimiterspace} D} \ll 1, \\ 
 F_{\rm in}^0 {\rm\;\;\;\;\;\;\;\;\;\;\; for \;\;\;\; }{U \mathord{\left/
 {\vphantom {U D}} \right.
 \kern-\nulldelimiterspace} D} \gg 1. \\ 
 \end{array} \right.
\label{42}
\ee
It can be seen (Eq. (\ref{41})) that the inter-site exchange interaction, $J$, in an effective field (\ref{41}) is opposing antiferromagnetism. This is contrary to the case of 
ferromagnetism (see Ref. \cite{10}). 

Using only the first order approximation for the self-energies (Eq. (\ref{39})) in Eq. (\ref{38}) we obtain the following expressions for electron numbers:
\be
n_\alpha ^{ \pm \sigma }  = n_\beta ^{ \mp \sigma }  = \frac{1}{{2N}}\sum\limits_k {\left[ {P_k^{ \pm \sigma } f(E_k ) + P_k^{ \mp \sigma } f( - E_k )} \right]}, 
\label{43}
\ee
where $f(E_k )$ is the Fermi function
\be
f(E_k ) = \frac{1}{{1 + \exp \left[ {\beta \left( {E_k  - \mu } \right)} \right]}},
\label{44}
\ee
\be
P_k^{ \pm \sigma }  = \frac{{\varepsilon _k S_k^{ \pm \sigma } }}{{E_k }}
\label{45}
\ee
is the occupation probability of state $(k, \pm \sigma )$,
\be
E_k  = \sqrt {\varepsilon _k^2  + \Delta ^2 } 
\label{46}
\ee
and
\be
\varepsilon _k  = \varepsilon _k^0 L_{\rm AF},\;\;\;\;\;\; S_k^{ \pm \sigma }  = \sqrt {\frac{{E_k  \mp \Delta }}{{E_k  \pm \Delta }}}. 
\label{47}
\ee
The previous results for the antiferromagnetism (see Ref. \cite{13}) are arrived at without the effect of bandwidth modification. In the present case, the kinetic energy still sums up 
to the previously used standard form (see Ref. \cite{14}):
\be
\sum\limits_k {\left( {\varepsilon _k S_k^\sigma   + \varepsilon _k S_k^{ - \sigma } } \right)}  = \sum\limits_k {\varepsilon _k \left( {S_k^\sigma   + S_k^{ - \sigma } } \right)}  = 
2\sum\limits_k {E_k }, 
\label{48}
\ee
where now, in the expression for effective dispersion relation of the antiferromagnet; $E_k  = \sqrt {\varepsilon _k^2  + \Delta ^2 }$, the dispersion relation $\varepsilon _k$ was 
replaced by $\varepsilon _k  = \varepsilon _k^0 L_{\rm AF} \left( {n,m} \right)$.

Using definitions of the average electron numbers $n_\gamma ^{ \pm \sigma } \,\left( {\gamma  = \alpha ,\beta } \right)$ (see Eq. (\ref{22})), and expression (\ref{43}) we obtain 
that in the mean-field approximation, the antiferromagnetic moment per atom (in Bohr's magnetons) is given by the following expression:
\be
m = n_\alpha ^\sigma   - n_\alpha ^{ - \sigma }  = \frac{1}{{2N}}\sum\limits_k {\left( {P_k^{ + \sigma }  - P_k^{ - \sigma } } \right)\left[ {f(E_k ) - f( - E_k )} \right]}. 
\label{49}
\ee
The chemical potential $\mu$ is determined from carrier concentration $n$ on the base of the equation, which is also coming from the Eq. (\ref{43}):
\be
n = n_\alpha ^\sigma   + n_\alpha ^{ - \sigma }  = \frac{1}{{2N}}\sum\limits_k {\left( {P_k^{ + \sigma }  + P_k^{ - \sigma } } \right)\left[ {f(E_k ) + f( - E_k )} \right]}. 
\label{50}
\ee 
Inserting Eq. (\ref{45}) to Eq. (\ref{49}) and using relation (\ref{40}) we obtain
\be
1 =  - F_{\rm AF} \sum\limits_k {\frac{1}{{2E_k }}\left[ {f(E_k ) - f( - E_k )} \right]}. 
\label{51}
\ee
 At the transition from AF to normal state condition (\ref{51}) takes on the form
 \be
1 =  - F_{\rm AF}^{\rm cr} \sum\limits_k {\frac{{\left[ {f(\varepsilon _k ) - f( - \varepsilon _k )} \right]}}{{2\varepsilon _k }}}. 
\label{52}
\ee
The numerical analysis will be based on Eqs (\ref{52}) and (\ref{50}). It will show how the critical exchange field $F_{\rm AF}^{\rm cr}$ depends on electron concentration. The 
analysis of Eqs (\ref{51}) and (\ref{50}) will give the magnetization dependence on the temperature for different values of the inter-site interactions.

\vskip0.5cm 
\noindent {\Large {\bf 5. Numerical examples}} 
\vskip0.5cm 

To illustrate the numerical results for the antiferromagnetic ordering, the Eqs (\ref{51}), (\ref{52}) and (\ref{50}) will be used in the integral form:
\be
1 =  - F_{\rm AF} \int\limits_{ - D}^D {\rho _0 \frac{{\left[ {f(E(\varepsilon )) - f( - E(\varepsilon ))} \right]}}{{2E(\varepsilon )}}d\varepsilon }, 
\label{53}
\ee 
\be
1 =  - F^{\rm cr}_{\rm AF} \int\limits_{ - D}^D {\rho _0 \frac{{\left[ {f(\varepsilon ) - f( - \varepsilon )} \right]}}{{2\varepsilon }}d\varepsilon } 
\label{54}
\ee
and
\be
n = \int\limits_{ - D}^D {\rho _0 \frac{{\left[ {f(E(\varepsilon )) + f( - E(\varepsilon ))} \right]}}{2}d\varepsilon }, 
\label{55}
\ee
where variable $\varepsilon$ replaced the effective dispersion relation $\varepsilon _k$, $E(\varepsilon ) = \sqrt {\varepsilon ^2  + \Delta ^2 }$ and $\rho_0$ is the constant DOS. 
This DOS for the weak correlation $\left( {U/D \ll 1} \right)$ is given by
\be
\rho _0  = \left\{ \begin{array}{l}
 \frac{1}{{D_0 L_{\rm AF} }}{\rm  \;\;\;\;\;\;\;\; for \;\;\;\; } - D_0 L_{\rm AF}  \le \varepsilon  \le D_0 L_{\rm AF}  \\ 
 {\rm  }0{\rm \;\;\;\;\;\;\;\;\;\;\;\;\;\;\;  for \; other \; }\varepsilon  \\ 
 \end{array} \right.
\label{56}
\ee
and for the strong correlation $\left( {U/D \gg 1} \right)$
\be
\rho _0  = \left\{ \begin{array}{l}
 \frac{{\sqrt {1 - {\textstyle{n / 2}}} }}{{D_0 L_{\rm AF} }}{\rm \;\;\;\;\;\;\;\;  for  \;\;\;\;} - \sqrt {1 - {\textstyle{n / 2}}} D_0 L_{\rm AF}  \le \varepsilon  \le \sqrt {1 - 
{\textstyle{n / 2}}} D_0 L_{\rm AF},  \\ 
 \frac{{\sqrt {{\textstyle{n / 2}}} }}{{D_0 L_{\rm AF} }}{\rm \;\;\;\;\;\;\;\;\;\;\;\, for  \;\;\;\;} - \sqrt {{\textstyle{n / 2}}} D_0 L_{\rm AF}  \le \varepsilon  - U \le \sqrt 
{{\textstyle{n / 2}}} D_0 L_{\rm AF},  \\ 
 0{\rm \;\;\;\;\;\;\;\;\;\;\;\;\;\;\;\;\;\;\; for \; other \;}\varepsilon.  \\ 
 \end{array} \right.
\label{57}
\ee
The effective bandwidth $D$, which appears as the integral boundaries in Eqs (\ref{53})--(\ref{55}), also depends on the Coulomb correlation $U$ and for the weak correlation 
$\left( {U/D \ll 1} \right)$ is equal to
\be
D = D_0 L_{\rm AF}, 
\label{58}
\ee
while for the strong correlation $\left( {U/D \gg 1} \right)$
\be
D = \left\{ \begin{array}{l}
 \sqrt {1 - {\textstyle{n / 2}}} D_0 L_{\rm AF} {\rm \;\;\;\;\;\;\;\; for \;lower \;Hubbard \;sub-band}, \\ 
 \sqrt {{\textstyle{n / 2}}} D_0 L_{\rm AF} {\rm \;\;\;\;\;\;\;\;\;\;\;\;\;\,\,  for \;upper \;Hubbard \;{\rm sub-band}}. \\ 
 \end{array} \right.
\label{59}
\ee
The parameter $I_{\rm AF}$, which appears in the expression (\ref{19}) for the bandwidth factor $L_{\rm AF}$, according to Eq. (\ref{5}) depends also on the Coulomb's 
correlation $U$ throughout quantity $n_i^{t\sigma }$. For the weak correlation $\left( {U/D \ll 1} \right)$ we have $n_i^{t\sigma }=1$ in Eq. (\ref{5}). Using additional values of 
$n_{\alpha \left( \beta  \right)}^\sigma$ from Eq. (\ref{22}) we obtain
\be
I_{\rm AF}  = \frac{{2n - n^2  - m^2 }}{{4(1 - m^2 )}}.
\label{60}
\ee
For the strong correlation $\left( {U/D \gg 1} \right)$, when $n_i^{t\sigma }  = 1 - n_i^{ - \sigma }$ (for the chemical level in the lower sub-band), we obtain from Eq. (\ref{5}) that
\be
I_{\rm AF}  = \frac{{\left( {2n - n^2  - m^2 } \right)\left( {1 - n} \right)}}{{(2 - n)^2  - m^2 }},
\label{61}
\ee
when at $m \to 0$ brings the following expression:
\be
I_0  = \frac{{n\left( {1 - n} \right)}}{{2 - n}}.
\label{62}
\ee
This expression can be again calculated directly from the definition \cite{4,9} of the average bond occupation with spin $\sigma$ in the paramagnetic limit of the band being split 
by Hubbard correlation
\be
i_0  =  - \int\limits_{ - \infty }^\infty  {\rho _0 \frac{\varepsilon }{D}f(\varepsilon )d\varepsilon }, 
\label{63}
\ee
where the effective bandwidth $D$ is given by (\ref{59}) and $\rho _0$ is given by (\ref{57}).

We assume the zero absolute temperature, for which the analytical expressions are obtained from Eqs (\ref{54}) and (\ref{55}) for the critical value of effective exchange field 
$F^{\rm cr}_{\rm AF}$ in the function of electron concentration $n$. In case of the weak correlation $\left( {U/D \ll 1} \right)$ , the effective exchange field $F^{\rm cr}_{\rm AF}$ 
is given by
\be
F_{\rm AF}^{\rm cr}  = \frac{{D_0 }}{{\log \left| {1 - n} \right|}}L_{\rm AF}. 
\label{64}
\ee
For the strong correlation $\left( {U/D \gg 1} \right)$, the following expression is obtained:
\be
F_{\rm AF}^{\rm cr}  = \frac{{D_0 }}{{\sqrt {1 - {\textstyle{n \over 2}}} \log \left| {\frac{{2 - n}}{{3n - 2}}} \right|}}L_{\rm AF}. 
\label{65}
\ee
The critical value of the effective exchange field $F_{\rm AF}^{\rm cr}$ depends strongly on the electron concentration and on the parameter of effective bandwidth $L_{\rm 
AF}$. Quantity $F_{\rm AF}^{\rm cr}$ decreases with decreasing bandwidth. The components of the effective exchange field are the on-site and inter-site exchange interactions 
(see Eq. (\ref{41})). It is possible to notice that the on-site exchange interaction $F_{\rm in}$ favors antiferromagnetism, but the inter-site interaction $J$ opposes it.
\vskip1.5cm
\begin{figure}[b]
\epsfig{file=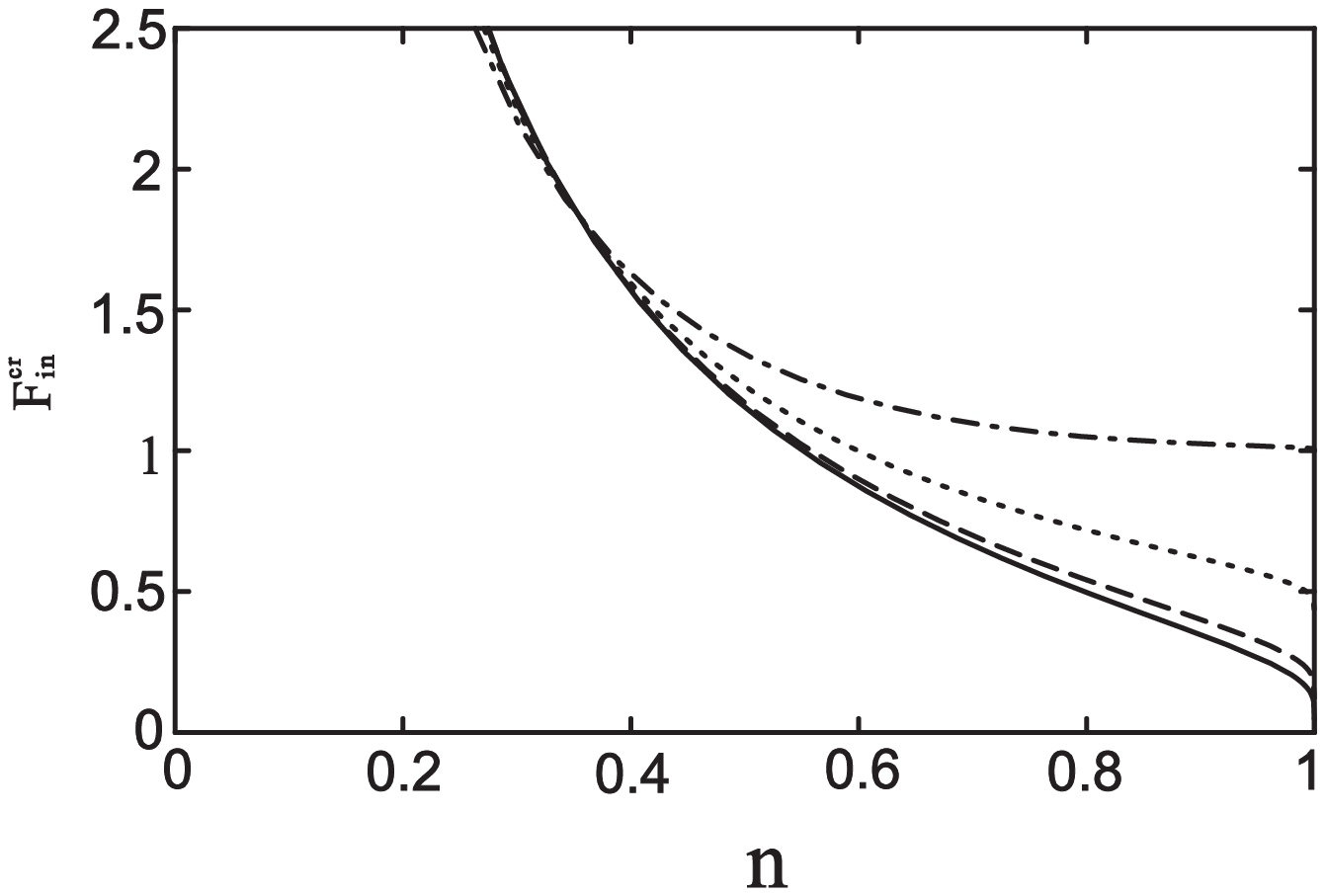,width=0.45\hsize}
 \hspace{0.05\hsize}
 \epsfig{file=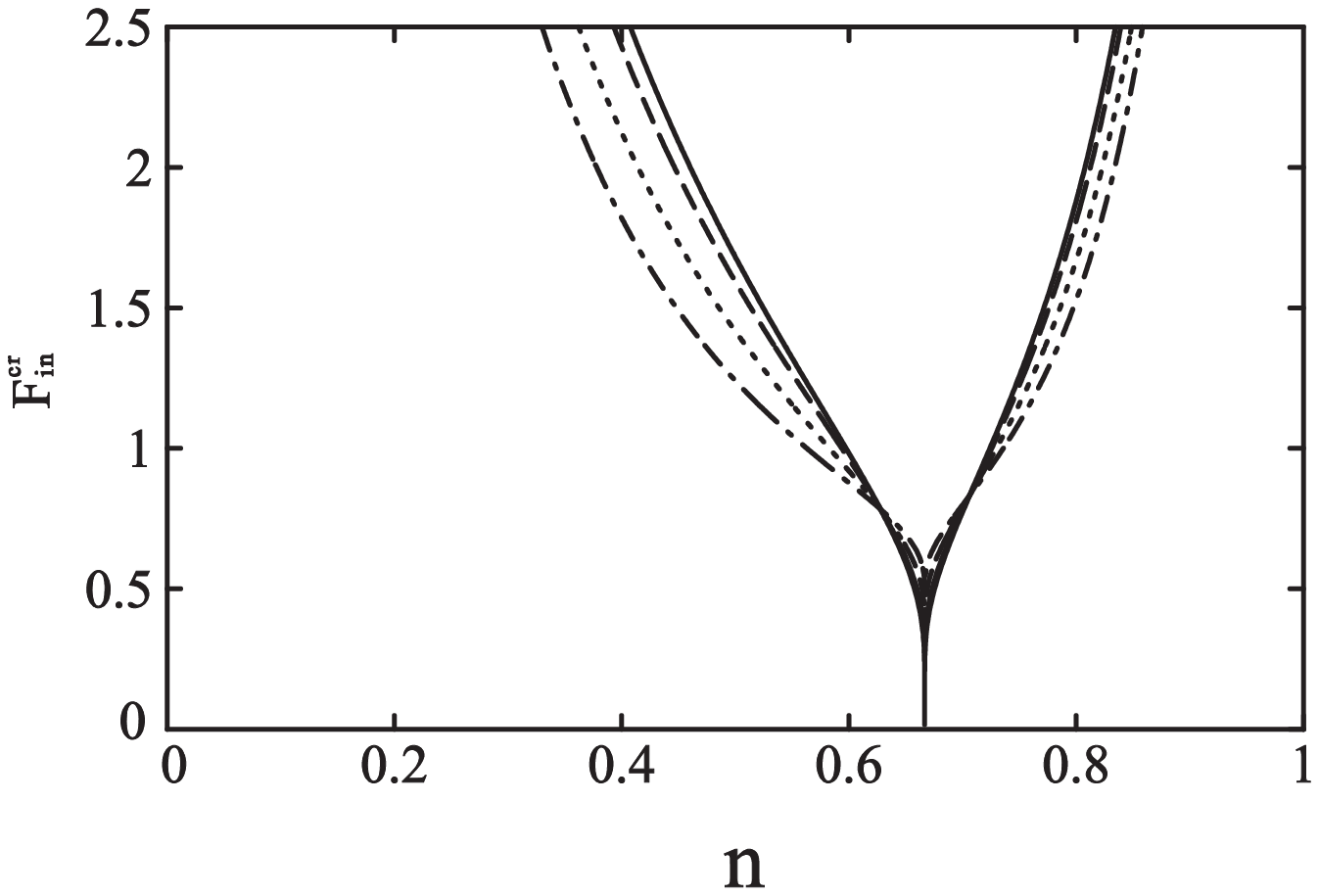,width=0.45\hsize}
 \par\vspace{1.5ex}\makebox[0.5\hsize]
    {\small FIG. 1a}\makebox[0.5\hsize]{\small FIG. 1b}
	\vskip0.5 cm
	\caption{ Dependence of the critical internal field $F^{\rm cr}_{\rm in}$ on electron concentration for different values of the inter-site exchange interaction $J$ in the case of 
weak (FIG. 1a) and strong (FIG. 1b) Coulomb correlation, at $t_0=0.1\,\rm eV,\, z=8,\, J'=J$ and $V=\Delta t=0;\, J=0$ --- solid line, $J=0.1t_0$ --- dashed line, $J=0.3t_0$ --- dotted 
line, $J=0.5t_0$ --- dotted-dashed line.}
\vskip0.5 cm
\end{figure}

Figs 1a and 1b show the dependence of the critical internal field $F_{\rm in}^{\rm cr}$ (given by Eq. (\ref{42})) on electron concentration for different values of the inter-site 
exchange interaction $J$. Additionally, the pair-hopping interaction $J'=J$ \cite{7} and $V=\Delta t=0$ is assumed. Analyzing the critical field $F_{\rm in}^{\rm cr}$ for the weak 
Coulomb correlation (Fig. 1a) one can see that close to half-filling the increase in the inter-site exchange interaction $J$ does not help antiferromagnetism. The effect is reversed for 
smaller concentrations. There is a competition here between two effects. The increase of interaction $J$ decreases the bandwidth $L_{\rm AF}$ (see Eq. (\ref{19})) which 
decreases $F_{\rm AF}^{\rm cr}$, but on the other side, it weakens the effective exchange interaction $F_{\rm AF}$ in accordance with Eq. (\ref{41}). Similar behavior is 
observed for the strong Coulomb correlation (see Fig. 1b) around the point of maximum in DOS which is shifted now due to the band splitting to the concentration of $n=2/3$.

\begin{figure}[t]
\epsfig{file=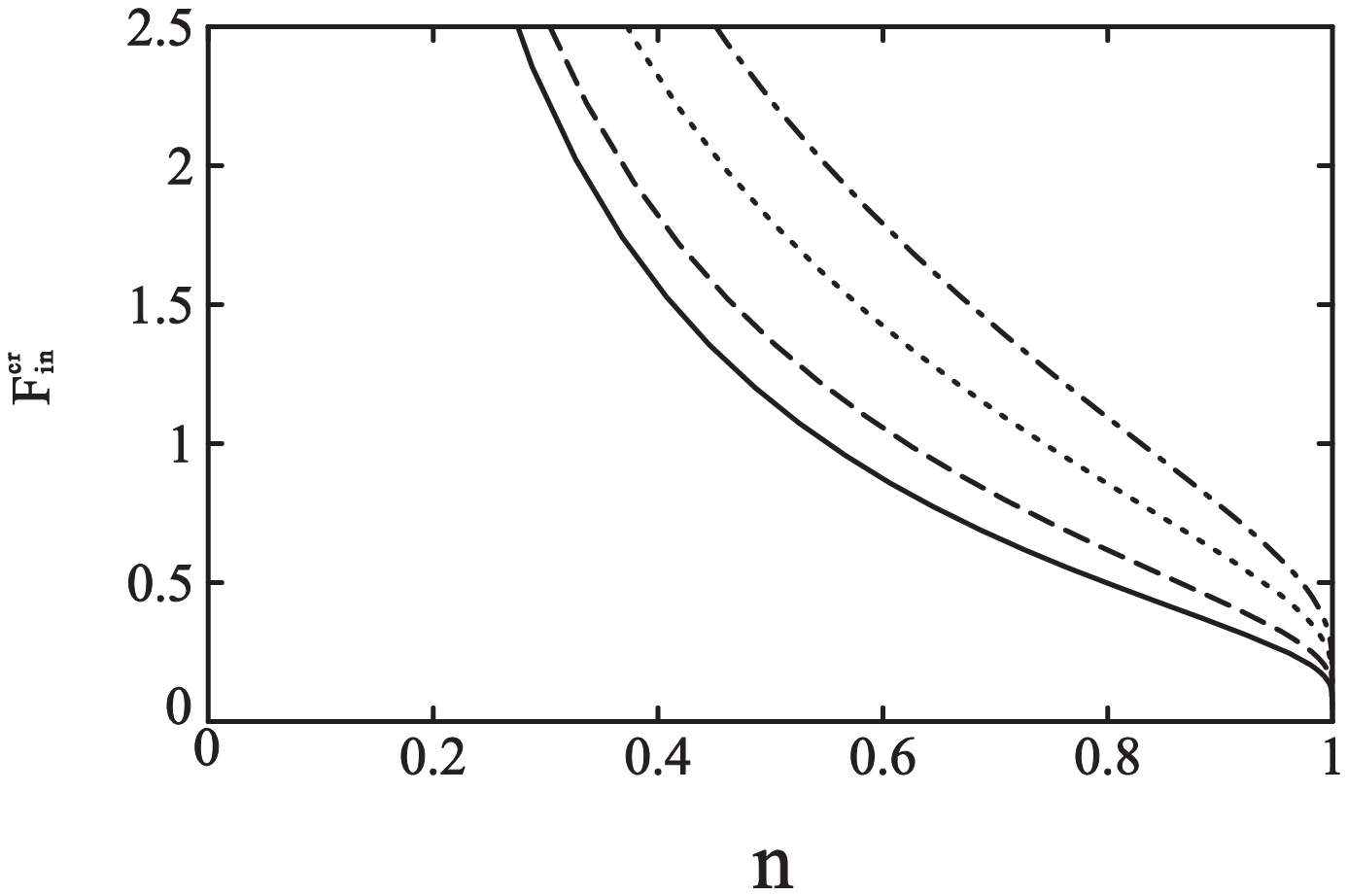,width=0.45\hsize}
 \hspace{0.05\hsize}
 \epsfig{file=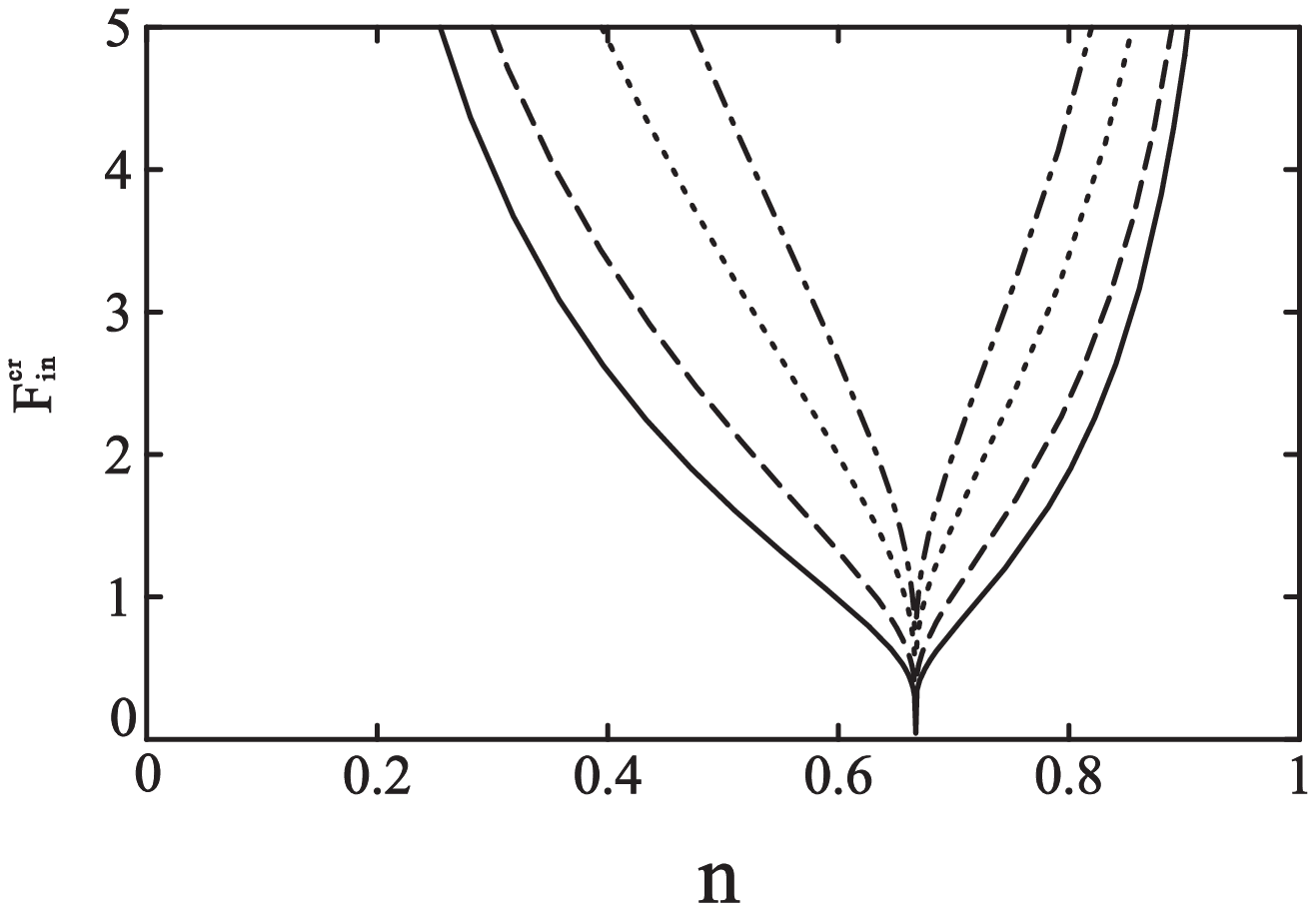,width=0.45\hsize}
 \par\vspace{1.5ex}\makebox[0.5\hsize]
    {\small FIG. 2a}\makebox[0.5\hsize]{\small FIG. 2b}
	\vskip0.5 cm
	\caption{ Dependence of the critical internal field $F^{\rm cr}_{\rm in}$ on electron concentration for different values of the density-density interaction $V$ in the case of weak 
(FIG. 2a) and strong (FIG. 2b) Coulomb correlation at $t_0=0.1\,\rm eV,\, z=8$ and $J=J'=\Delta t=0;\, V=0$ --- solid line, $V=1t_0$ --- dashed line, $V=3t_0$ --- dotted line, $V=5t_0$ 
--- dotted-dashed line.}
\vskip0.5 cm
\end{figure}

Figs 2a and 2b present the dependence of the critical exchange field on the electron concentration for different values of the density-density interaction $V$. The other inter-site 
interactions are equal zero; $J=J'=\Delta t=0$. Interaction $V$ increases the bandwidth hence it increases the critical field $F_{\rm AF}^{\rm cr}$. This in turn decreases the 
concentration range of possible antiferromagnetic state.

\begin{figure}[t]
\epsfig{file=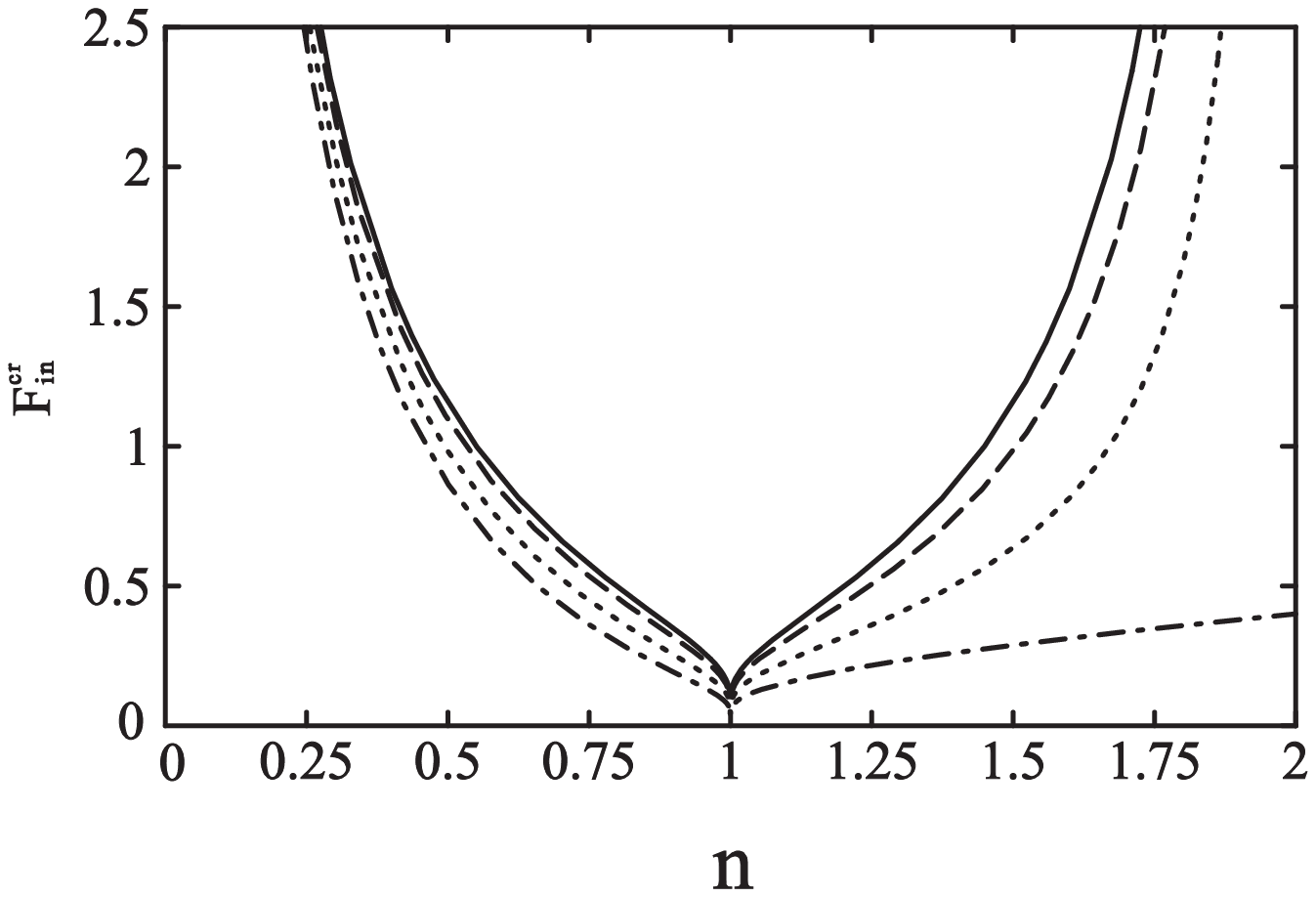,width=0.45\hsize}
 \hspace{0.05\hsize}
 \epsfig{file=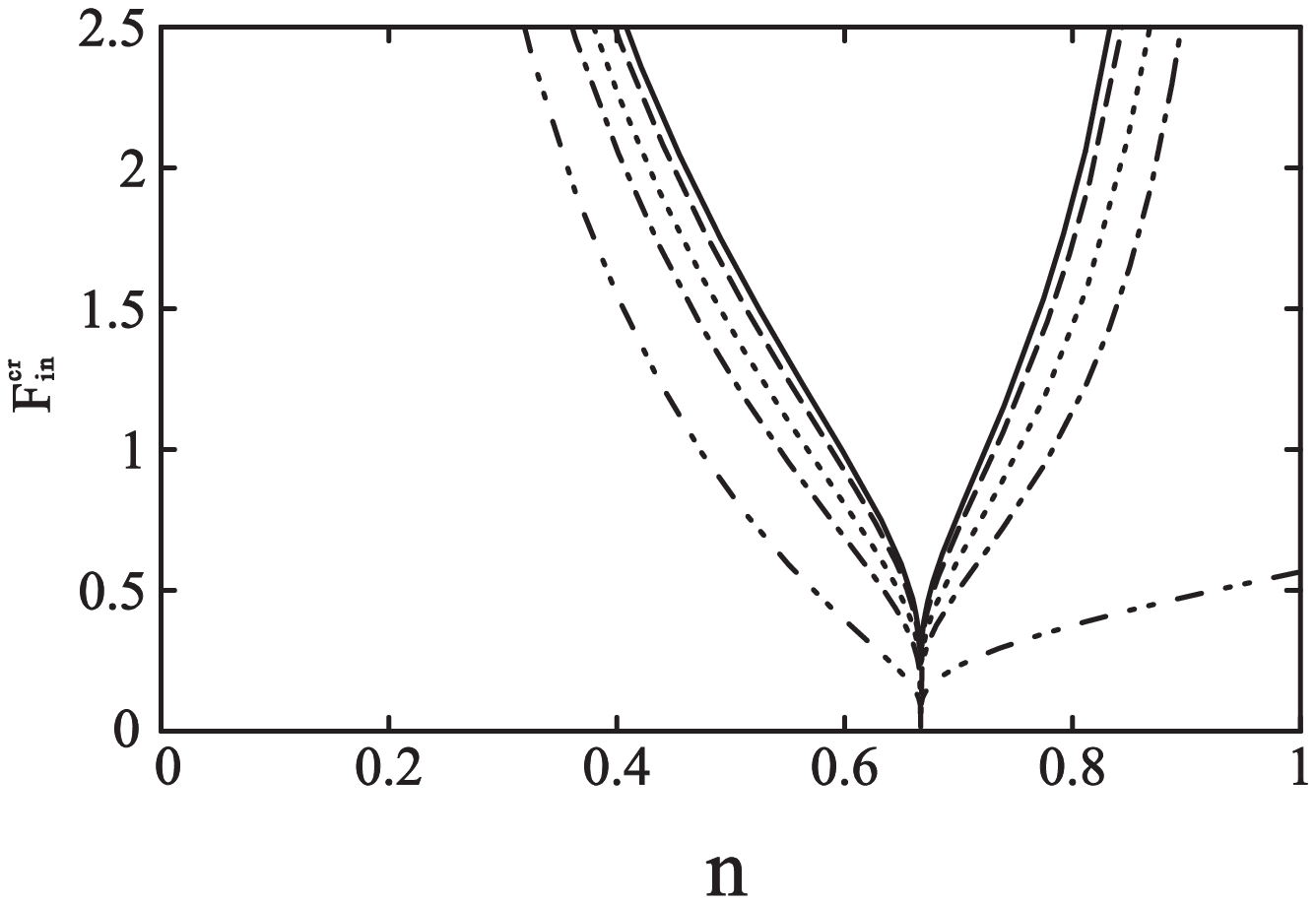,width=0.45\hsize}
 \par\vspace{1.5ex}\makebox[0.5\hsize]
    {\small FIG. 3a}\makebox[0.5\hsize]{\small FIG. 3b}
	\vskip0.5 cm
	\caption{Dependence of the critical internal field $F^{\rm cr}_{\rm in}$ on electron concentration for different values of the assisted hopping interaction  $\Delta t$ in the case 
of weak (FIG. 3a) and strong (FIG. 3b) Coulomb correlation at $t_0=0.1\,\rm eV,\, z=8$ and $J=J'=V=0;\, \Delta t=0$ --- solid line, $\Delta t=0.1t_0$ --- dashed line, $\Delta t=0.3t_0$ 
--- dotted line, $\Delta t=0.5t_0$ --- dotted-dashed line.}
\vskip0.5 cm
\end{figure}

The next interaction, which will be analyzed, is the assisted hopping interaction $\Delta t$. According to Eq. (\ref{19}), the increase of $\Delta t$ decreases the bandwidth, which 
decreases the total critical exchange field $F_{\rm AF}^{\rm cr}$. Figs 3a and 3b show the dependence of the internal critical field $F_{\rm in}^{\rm cr}$ on electron concentration 
for the weak and strong Coulomb correlation respectively, for different values of the assisted hopping interaction $\Delta t$. The other inter-site interactions are equal to zero; 
$J=J'=V=0$. The change in the bandwidth $L_{\rm AF}$ caused by assisted hopping interaction $\Delta t$, unlike the changes from all other inter-site interactions, is not 
symmetrical with respect to the half-filling. In effect, the dependence $F_{\rm in}^{\rm cr}(n)$ is not symmetrical with respect to the half-filling. Nonetheless, for $U = \infty$ we 
can draw curves in Fig. 3b only to the concentration of $n=1$. The electronic states for $n>1$ are not available for energetic reasons.

\begin{figure}[t]
\begin{center}
\epsfig{file=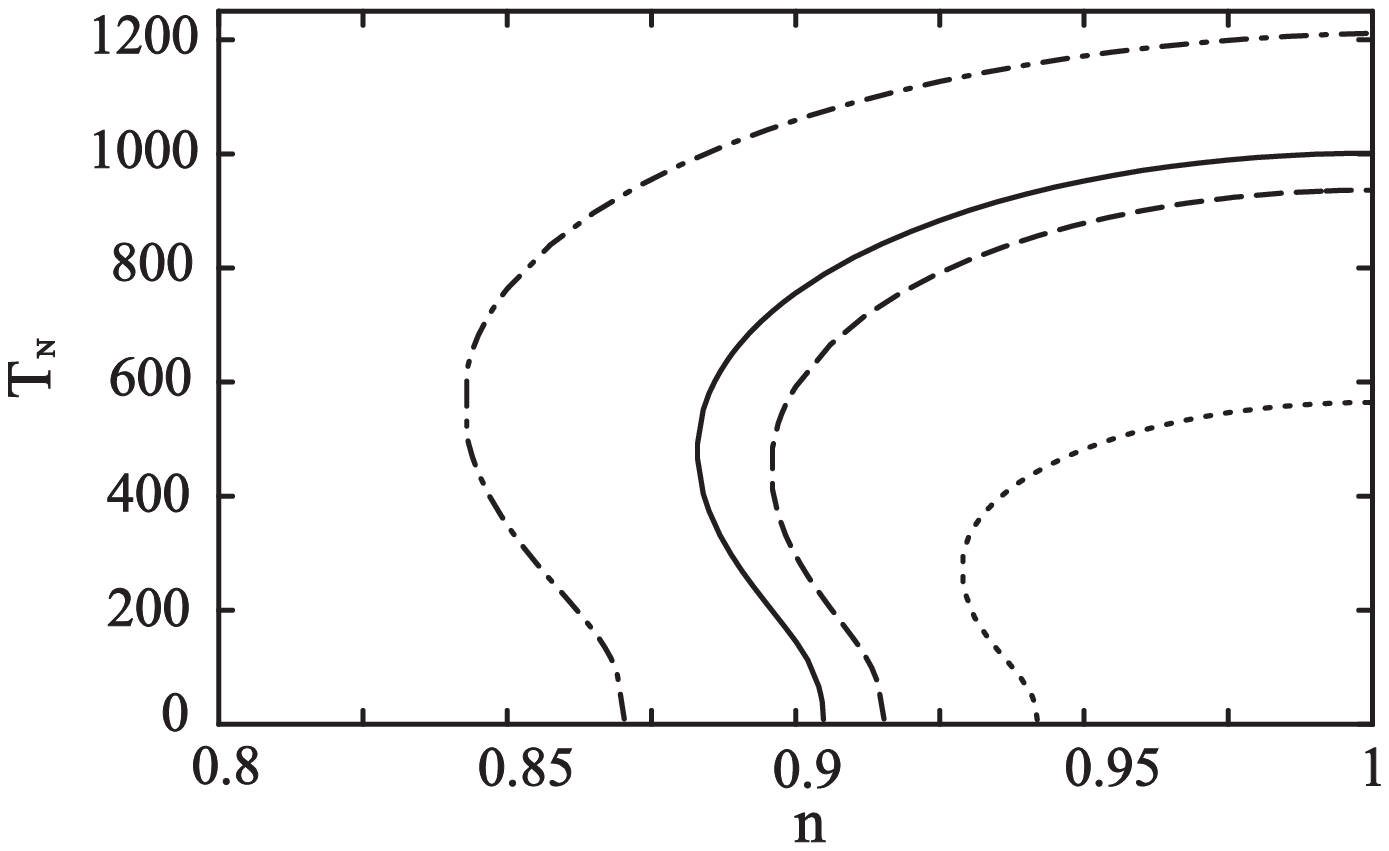,width=0.45\hsize}
 \par\vspace{1.5ex}\makebox[0.5\hsize]
    {\small FIG. 4}
	\end{center}
	\vskip0.5 cm
	\caption{Dependence of the N\'eel's temperature $(T_{\rm N})$ on electron concentration for different values of the inter-site interactions in the case of weak Coulomb 
correlation at $t_0=0.1\,\rm eV,\, z=8$ and $F_{\rm in}=0.34 \,\rm eV$; $J=J'=V=\Delta t=0$ --- solid line, $J=J'=\Delta t=0$ and $V=0.2t_0$ --- dashed line, $V=\Delta t=0$ and 
$J=J'=0.1t_0$ --- dotted line, $J=J'=V=0$ and $\Delta t=0.15t_0$ --- dotted-dashed line.}
\vskip0.5 cm
\end{figure}

Fig 4 presents the dependence of N\'eel's temperature $T_{\rm N}$, on electron concentration for different values of the inter-site interactions for the weak Coulomb correlation. 
All the curves are for the same internal exchange field, which is $F_{\rm in}=0.34 \,\rm eV$. The curves with nonzero inter-site exchange interaction $J$ and nonzero 
density-density interaction $V$, have smaller magnetization and the N\'eel's temperature. For half-filling $(n=1)$ they are $564\, \rm K$ and $936\,\rm K$ respectively. Without any 
inter-site interaction the N\'eel's temperature is higher, at half-filling it is $T_{\rm N}=1000\,\rm K$. The inter-site exchange interaction $J$ decreases the critical temperature by 
decreasing the effective exchange interaction $F_{\rm AF}$ (see Eq. (\ref{41})). This effect is stronger than the decrease of the bandwidth due to $J$, which increases $T_{\rm 
N}$. The density-density interaction $V$ decreases the critical temperature by increasing the bandwidth. Only the assisted hopping interaction $\Delta t$ increases the N\'eel's 
temperature (to $1211\,\rm K$ at half-filling) by decreasing the bandwidth.

\begin{figure}[t]
\begin{center}
\epsfig{file=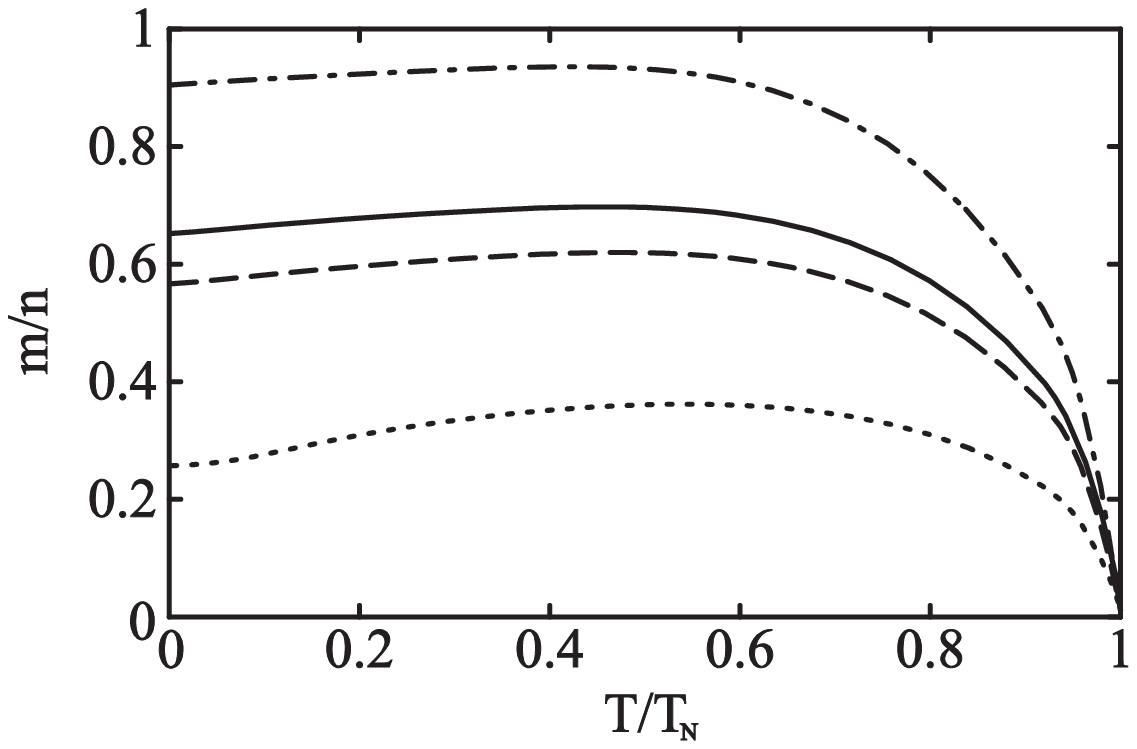,width=0.45\hsize}
 \par\vspace{1.5ex}\makebox[0.5\hsize]
    {\small FIG. 5}
	\end{center}
	\vskip0.5 cm
	\caption{Magnetization versus temperature for band filling $n=0.95, \,t_0=0.1\,\rm eV, \,z=8, \,F_{\rm in}=0.34\, \rm eV$ and for various values of the inter-site interaction; 
$J=J'=V=\Delta t=0$ --- solid line, $J=J'=\Delta t=0$ and $V=0.2t_0$ --- dashed line, $V=\Delta t=0$ and $J=J'=0.1t_0$ --- dotted line, $J=J'=V=0$ and $\Delta t=0.15t_0$ --- 
dotted-dashed line.}
\vskip0.5 cm
\end{figure}

Figure 5 shows the magnetization versus temperature for $n=0.95$ and various values of the inter-site interaction. Again, all the curves are for the same internal exchange field 
$F_{\rm in}=0.34 \,\rm eV$. Similarly, as in Fig. 4 at constant value of internal field, the highest magnetization is obtained for nonzero assisted hopping interaction $\Delta t$.

\begin{figure}[t]
\begin{center}
\epsfig{file=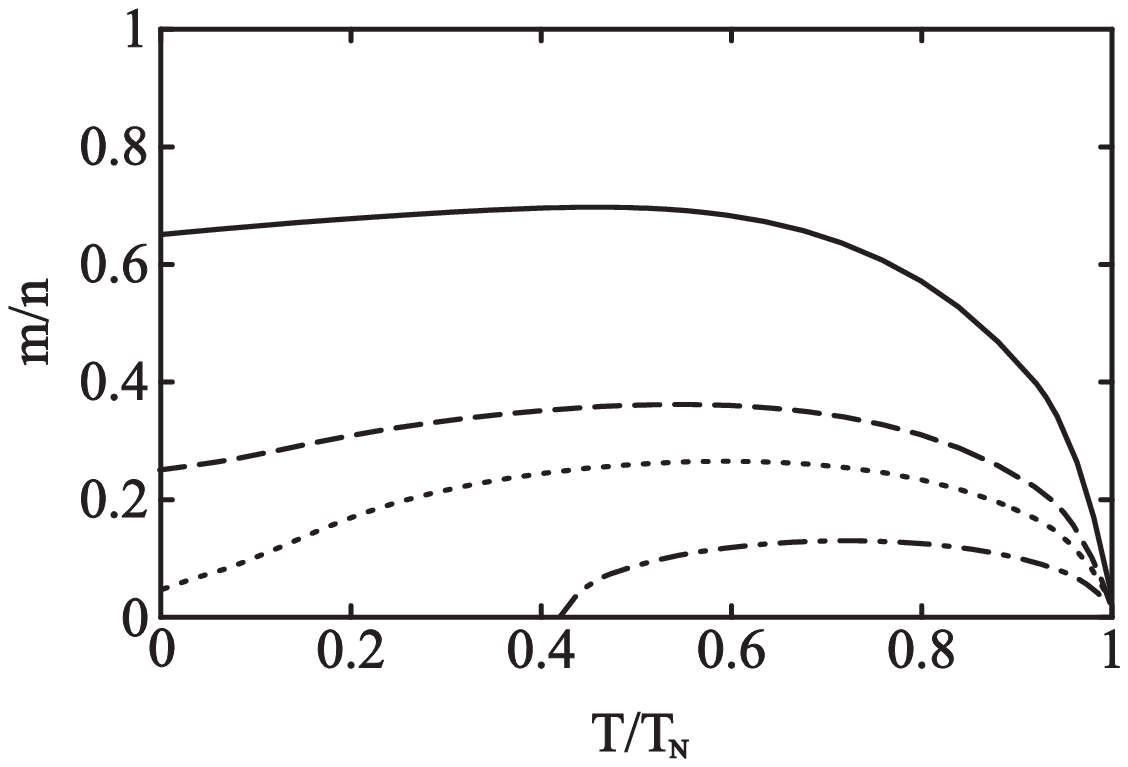,width=0.45\hsize}
 \par\vspace{1.5ex}\makebox[0.5\hsize]
    {\small FIG. 6}
	\end{center}
	\vskip0.5 cm
	\caption{Magnetization versus temperature for band filling $n=0.95, \,t_0=0.1\,\rm eV, \,z=8, \,F_{\rm in}=0.34\, \rm eV$ and for various values of the inter-site exchange 
interaction $J$ at $V=\Delta t=0$ and $J'=J$; $J=0$ --- solid line, $J=0.1t_0$ --- dashed line, $J=0.12t_0$ --- dotted line, $J=0.14t_0$ --- dotted-dashed line.}
\vskip0.5 cm
\end{figure}

Figure 6 shows the magnetization versus temperature for $n=0.95$ and various values of the inter-site exchange interaction $J$. The value of internal exchange field is $F_{\rm 
in}=0.34 \,\rm eV$. Magnetization is decreasing with increasing inter-site exchange interaction. For $J \approx 0.122t_0$ the magnetization approaches zero as $T\to 0$, but for 
higher temperatures the reentrant antiferromagnetism occurs. For $J>0.15t_0$ the antiferromagnetic state at our values of $n=0.95$ and $F_{\rm in}=0.34 \,\rm eV$ does not exist at 
any temperatures.  

\vskip0.5cm 
\noindent {\Large {\bf 6. Conclusions}} 
\vskip0.5cm

The most essential results of this paper will be recapitulated now. 

Introducing the inter-site interactions does not change qualitatively the results obtained much earlier (see \cite{13}) for the antiferromagnetic ordering coming from the Coulomb 
correlation. In the case of the weak Coulomb correlation, there is a minimum of the critical exchange field at the half-filling. In the case of the strong Coulomb correlation, the 
minimum of these curves shifts close to the half-filling of the lower sub-band $(n=2/3)$ and the half-filling of the upper sub-band $(n=4/3)$.

The inter-site nearest-neighbor interactions lower the critical exchange field by the ratio of $L_{\rm AF}  = \frac{D}{{D_0 }} = 1 - n\frac{{\Delta t}}{{t_0 }} + \frac{1}{{t_0 
}}\left( { - 2J - J' + V} \right)I_{\rm AF}$. At the same time, the inter-site exchange interaction, $J$, increases the necessary critical exchange field by directly opposing it in the 
Weiss field; $F_{\rm AF}^{\rm cr}  = F_{\rm in}^{\rm cr}  - zJ$. The net effect of these interactions was illustrated in Figs \mbox{1--3}.

It has already been established that the superconducting cuprates have the superconducting band being split by the relatively strong Coulomb repulsion into two sub-bands, 
each with the maximum capacity of one electron. The new approach developed here allows for the antiferromagnetic instability in the strong correlation split band limit to occur at 
the half-filling, see Fig. 3b, when the assisted hopping interaction $\Delta t \approx t_0$, which reduces the bandwidth to zero, $L_{\rm AF}\approx 0$, at this concentration. At 
$J=J'=V=0$ we obtain from Eq. (\ref{65}) the following formula:
\be
F_{\rm AF}^{\rm cr}  = \frac{{D_0 }}{{\sqrt {1 - {\textstyle{n / 2}}} \log \left| {\frac{{2 - n}}{{3n - 2}}} \right|}}\left( {1 - \frac{{\Delta t}}{{t_0 }}} \right).
\label{66}
\ee
The other inter-site interactions $(J, J', V)$ do not reduce the bandwidth at $n=1$, since $I_{\rm AF}(n=1)=0$ (see Eq. (\ref{61})) in the case of the strong Coulomb correlation.

The reduction of the bandwidth to zero for strong correlation at half-filling in the purely itinerant model can also be caused by the limitation of electron jumps at this 
concentration, which according to Ref. \cite{15} decreases the bandwidth $t = t_0 \sqrt {\left( {1 - n} \right)/\left( {1 - n/2} \right)}$. At close inspection of the expression for 
$F^{\rm cr}_{\rm AF}$ coming from Eq. (\ref{54}) at constant DOS and zero absolute temperature, the following formula is arrived at
\be
F_{\rm AF}^{\rm cr}  = \frac{{D_0 L_{\rm AF} \sqrt {\left( {1 - n} \right)/\left( {1 - n/2} \right)} }}{{\log (1 - n)}},
\label{67}
\ee 
with $L_{\rm AF}$ given by Eq. (\ref{19}). This expression shows that $F_{\rm AF}^{\rm cr}$ coming from this effect also tends to zero at the half-filling.
 
In conclusion the antiferromagnetism at half-filling in the high correlation case can appear when the bandwidth goes to zero at this concentration. It can go to zero either as 
$1-n(\Delta t/t_0)$, due to the presence of the assisted hopping (or some other interaction, whose strength is proportional to $n$) or as $\sqrt {1 - n}$, due to the hopping 
limitation in the strong correlation limit.

Both these situations can describe the high temperature superconducting cuprates, where at the half-filling and the strong Coulomb correlation there is, initially, the 
antiferromagnetic order. The superconductivity will appear upon doping only away from the half-filling at the electron concentration $n \approx 0.95 - 0.8$ (see e.g. Ref. \cite{16}).

\newpage
\noindent {\Large {\bf Appendix}} 
\vskip0.5cm 
The stochastic potential in Hamiltonian (\ref{8}) can be expressed as
$$
\tilde \varepsilon  = (J + J')\overline {c_{j - \sigma }^ +  c_{i - \sigma } }  + (J - V)\overline {c_{j\sigma }^ +  c_{i\sigma } }  + \Delta t\left( {\overline {n_{i - \sigma } }  + \overline 
{n_{j - \sigma } } } \right). \eqno (A1)
$$
To arrive at this equation the stochastic value replaced each operator product in the square bracket of Hamiltonian (\ref{3}). For example, the operator product; $c_{j\sigma }^ +  
c_{i\sigma }$ was replaced by the stochastic value $\overline {c_{j\sigma }^ +  c_{i\sigma } }$. It takes the values 1 or 0. The probability of value 1, $I_{\sigma}$, is the 
probability of electron with spin $\sigma$ hopping from the $i$ to the $j$ lattice site and back. The probability of the $i$i to $j$ lattice site hopping is given by the product of 
probabilities that there is an electron with spin $\sigma$ on the $i$ site and that the $j$ site has empty states; $n_i^\sigma  (n_j^{t\sigma }  - n_j^\sigma  )/\left( {n_i^\sigma   + 
n_j^{t\sigma }  - n_j^\sigma  } \right)$. The probability of forward and backward hopping is given by $I_{\sigma}$ from Eq. (\ref{5}) in the paper.

According to this idea the stochastic potential $\tilde \varepsilon$ from Eq. (A1) takes on one of the following values with corresponding probabilities
$$
\tilde \varepsilon  = \left\{ {\begin{array}{*{20}c}
   {\varepsilon _1  = J - V} \hfill & {P_1^\sigma   = I_\sigma  (1 - I_{ - \sigma } )(1 - n_i^{ - \sigma } )(1 - n_j^{ - \sigma } ),} \hfill  \\
   {\varepsilon _2  = J + J'} \hfill & {P_2^\sigma   = I_{ - \sigma } (1 - I_\sigma  )(1 - n_i^{ - \sigma } )(1 - n_j^{ - \sigma } ),} \hfill  \\
   {\varepsilon _3  = \Delta t} \hfill & {P_3^\sigma   = n_i^{ - \sigma } (1 - I_\sigma  )(1 - I_{ - \sigma } )(1 - n_j^{ - \sigma } ),} \hfill  \\
   {\varepsilon _4  = \Delta t} \hfill & {P_4^\sigma   = n_j^{ - \sigma } (1 - I_\sigma  )(1 - I_{ - \sigma } )(1 - n_i^{ - \sigma } ),} \hfill  \\
   {\varepsilon _5  = 2J + J' - V} \hfill & {P_5^\sigma   = I_\sigma  I_{ - \sigma } (1 - n_j^{ - \sigma } )(1 - n_i^{ - \sigma } ),} \hfill  \\
   {\varepsilon _6  = J - V + \Delta t} \hfill & {P_6^\sigma   = I_\sigma  n_i^{ - \sigma } (1 - I_{ - \sigma } )(1 - n_j^{ - \sigma } ),} \hfill  \\
   {\varepsilon _7  = J - V + \Delta t} \hfill & {P_7^\sigma   = I_\sigma  n_j^{ - \sigma } (1 - I_{ - \sigma } )(1 - n_i^{ - \sigma } ),} \hfill  \\
   {\varepsilon _8  = J + J' + \Delta t} \hfill & {P_8^\sigma   = I_{ - \sigma } n_i^{ - \sigma } (1 - I_\sigma  )(1 - n_j^{ - \sigma } ),} \hfill  \\
   {\varepsilon _9  = J + J' + \Delta t} \hfill & {P_9^\sigma   = I_{ - \sigma } n_j^{ - \sigma } (1 - I_\sigma  )(1 - n_i^{ - \sigma } ),} \hfill  \\
   {\varepsilon _{10}  = 2\Delta t} \hfill & {P_{10}^\sigma   = n_j^{ - \sigma } n_i^{ - \sigma } (1 - I_{ - \sigma } )(1 - I_\sigma  ),} \hfill  \\
   {\varepsilon _{11}  = 2J + J' - V + \Delta t} \hfill & {P_{11}^\sigma   = I_\sigma  I_{ - \sigma } n_i^{ - \sigma } (1 - n_j^{ - \sigma } ),} \hfill  \\
   {\varepsilon _{12}  = 2J + J' - V + \Delta t} \hfill & {P_{12}^\sigma   = I_\sigma  I_{ - \sigma } n_j^{ - \sigma } (1 - n_i^{ - \sigma } ),} \hfill  \\
   {\varepsilon _{13}  = J + J' + 2\Delta t} \hfill & {P_{13}^\sigma   = I_{ - \sigma } n_i^{ - \sigma } n_j^{ - \sigma } (1 - I_\sigma  ),} \hfill  \\
   {\varepsilon _{14}  = J - V + 2\Delta t} \hfill & {P_{14}^\sigma   = I_\sigma  n_i^{ - \sigma } n_j^{ - \sigma } (1 - I_{ - \sigma } ),} \hfill  \\
   {\varepsilon _{15}  = 2J + J' - V + 2\Delta t} \hfill & {P_{15}^\sigma   = I_\sigma  I_{ - \sigma } n_i^{ - \sigma } n_j^{ - \sigma } ,} \hfill  \\
   {\varepsilon _{16}  = 0} \hfill & {P_{16}^\sigma   = (1 - I_\sigma  )(1 - I_{ - \sigma } )(1 - n_i^{ - \sigma } )(1 - n_j^{ - \sigma } ).} \hfill  \\
\end{array}} \right. \eqno(A2)
$$

\end{document}